\documentclass[9pt,english]{IEEEtran}
\usepackage[T1]{fontenc}
\usepackage[latin9]{inputenc}
\usepackage{geometry}
\usepackage{float}
\usepackage{amsmath}
\usepackage{amssymb}
\usepackage{graphicx}
\usepackage{bm}

\makeatletter

\floatstyle{ruled}
\newfloat{algorithm}{tbp}{loa}
\providecommand{\algorithmname}{Algorithm}
\floatname{algorithm}{\protect\algorithmname}

\usepackage{todonotes}

\usepackage{spconf}
\usepackage[numbers,sort&compress]{natbib}

\name{Hengyan Liu$^{\star}$ , Wei Dai$^{\star}$, and Yuan Shen$^{\dagger}$   \thanks{This work is partially supported by UK RS International Exchanges 2017 Cost Share, UK EPSRC Grant EP/S026622/1 and EP/S026657/1 and the UK MOD UDRC in Signal Processing. } }  

\address{$^{\star}$ Electrical and Electronic Engineering, Imperial College London, UK   \\ $^{\dagger}$Department of Electronic Engineering, Tsinghua University, China}

\@ifundefined{showcaptionsetup}{}{%
 \PassOptionsToPackage{caption=false}{subfig}}
\usepackage{subfig}
\makeatother

\usepackage{babel}
\begin{document}
%\title{Joint Super-resolution for mixed LOS/OLOS/NLOS Localisation }
\title{Super-resolved Localisation without Identifying LoS/NLoS Paths}
\maketitle

\makeatletter
\renewcommand\footnoterule{%
  \kern-3\p@
  \hrule\@width.4\columnwidth
  \kern2.6\p@}
  \makeatother

\begin{abstract}

%\noindent 
 This paper focuses on the problem of localising a transmitting mobile station (MS) using multiple cooperative base stations (BSs). 
There are two technical difficulties: one is the data association between intermediate parameters and scatters; the other is to identify Line-of-Sight (LoS) and Non-Line-of-Sight (NLoS) paths. 
Our main contribution is a unified approach bypassing both data association and LoS/NLoS path identification. 
This is achieved by introducing virtual scatters and then a direct localisation formulation. 
Modern super-resolution technique is then adapted to cast the localisation problem as convex programming and solve it. 
Our approach addresses long-standing issues not completely solved in the literature, guarantees a global convergence, achieves good localisation accuracy and demonstrates its robustness to noise in our numerical tests.

\end{abstract}


\section{Introduction \label{sec:Intro}}
\noindent
This paper considers the problem of localising a transmitting mobile station (MS) based on measurements of several base stations (BSs). This problem has become vital in various applications such as the location based services and the Internet of things. 
In the literature, such a localisation is typically achieved by using the following two types of information. It has been assumed that BSs are equipped with multiple antennas and therefore BSs can estimate Direction of Arrival (DoA) of the received signals, and that localisation is performed in a cooperative setting and hence Time of Arrival (ToA)/Time Difference of Arrival (TDoA) information can be estimated and shared across BSs.

There are two technical difficulties in solving the localisation problem. The first one comes from data association between DoA/ToA information and locations of scatters. A typical approach is that BSs first estimate DoA/ToA and then infer MS location based on them. When there are multiple scatters in the scene, the task of associating DoA/ToA estimates to scatters is an NP-hard problem. In the literature, direct localisation \cite{VTC2008,shahmansoori2017position,garcia2017direct,gear2019maximum} has been developed to address this issue by considering the map  from locations directly to received signals and eliminating the intermediate variables DoA/ToA. Data association can be avoided. 

The other difficulty is a consequence of different impacts of Line-of-Sight (LoS) and Non-Line-of-Sight (NLoS) paths. Lots of efforts have been made to distinguish these two types of paths and study the localisation problem under different conditions. The identification of LoS/NLoS paths is typically achieved by constructing probabilistic models \cite{borras1998decision,Markov_4,Markov_42,cong2001non}, applying statistical estimation methods \cite{riba2004non, chan2006time}, and/or exploring the temporal correlations in the propagation environment \cite{wylie1996non}. However, mistaken identifications cannot be avoided completely. When they happen the localisation performance deteriorate.

Localisation mechanism varies based on the propagation environment. In the literature, typically three conditions are discussed: LoS condition where all paths are LoS; Obstructed-Line-of-Sight (OLoS) condition where all paths are NLoS; and NLoS condition where there exist both LoS and NLoS paths. Triangulation has been applied for localisation in all three conditions. Triangulation is straightforward in LoS condition. There are two approaches in NLoS condition. One is to identify LoS and NLoS paths and then only use LoS paths for triangulation \cite{chan2006time,riba2004non,cong2001non}. The other does not require LoS/NLoS path identification \cite{seow2008non,zhang2009combined,miao2007positioning}. Instead, geometric areas (typically circular areas) are obtained based on ToA/DoA estimation at BSs. The MS must lie in the intersection of these areas. This idea can be also applied to OLoS condition. However, in OLoS condition, the intersection is typically an area rather than a point. Extra assumptions are typically added to allow path Direction of Departure (DoD) information at BSs. Most of the above works rely on identification of LoS/NLoS paths and estimation of ToA/DoA information. 
 
Our main contribution is a unified approach that bypasses both data association and the middle step of identifying LoS/NLoS paths. This is achieved by three key elements. Firstly, virtual scatters are added to LoS paths so that all LoS paths can be treated mathematically in the same way as for NLoS paths. There is no need for LoS/NLoS path identification. Secondly, a direct localisation approach is developed where the intermediate variables DoA/ToA get eliminated. Different from \cite{VTC2008,shahmansoori2017position,garcia2017direct,gear2019maximum} where the idea of direct localisation has been applied, we formulate the localisation problem as a linear de-mixing problem so that super-resolution technique can be applied. Thirdly, modern super-resolution technique is adapted where a convex optimisation problem is formulated and solved. Our unified approach avoids the error propagation due to LoS/NLoS path identification, eliminates data association, guarantees a global convergence, achieves good localisation accuracy and demonstrates its robustness to noise in our numerical tests. 

\section{The Problem }
\label{sec:Problem}
\noindent
We focus on the following localisation problem by adopting a commonly assumed setup in the literature \cite{cong2001non,seow2008non,miao2007positioning,zhang2009combined,yin2013and}. We consider a wireless communication system where $J$ BSs cooperate to jointly estimate the location of one MS. The MS is assumed to transmit with an omnidirectional antenna and all BSs are equipped with a uniform linear antenna array (ULA) with $N_{R}$ antenna elements. The synchronisation between MS and BSs can be avoided by using round-trip ToA information, or using TDoA information with synchronised BSs \cite{cong2001non}.  It is typically assumed that via control signalling, the MS and the involved BSs are synchronised \cite{zhang2009combined,yin2013and} and the transmitted waveform from the MS is known to BSs.  Thus BSs can estimate both DoA and ToA information. In our signal model, we only consider either LoS or single-bounced signals from the MS to the BSs. This is motivated by the fact that signals scattered twice or more times typically suffer from great propagation losses and are thus less perceptible \cite{miao2007positioning}. 
It is noteworthy that the above setup is a simplification of actual systems. For example, the assumption of synchronisation and the complete discard of multiple-bounced signals may be problematic in practice. Nevertheless, the above setup is widely adopted in the literature \cite{cong2001non,seow2008non,miao2007positioning,zhang2009combined,yin2013and} for the purpose of highlighting the technical approach/idea without being drowned into great technical details. 

The received signal $\bm{y}_{j}\left(t\right) \in \mathbb{C}^{N_R}$ at the $j$-th BS is given by
\begin{equation}\label{eq:signal-model}
\bm{y}_{j}\left(t\right) = \sum_{k=1}^{K} \gamma_{j,k} x\left(t-\tau_{j,k}\right) \bm{a}\left(\theta_{j,k}\right), 
\end{equation}
where the number $K$ denotes the number of received paths at the BS, $\gamma_{j,k}\in \mathbb{C}$ is an unknown coefficient modelling the signal attenuation along the $k$-th path, $x(t)\in \mathbb{C}$ is the transmitted signal, $\tau_{j,k}>0$ is the delay experienced in the $k$-th path, $\theta_{j,k}$ is the  DoA of the $k$-th path, $\bm{a}(\theta) \in \mathbb{C}^{N_R}$ reflects the phase differences in the received signals due to DoA and takes the form of  
\begin{equation}
\bm{a} \left(\theta\right) = \left[ 1, e^{j\frac{2\pi}{\lambda}L\sin(\theta)},\cdots,e^{j\frac{2\pi}{\lambda}L\sin(\theta)\left(N_{R}-1\right)}\right]^{\mathrm{T}},
\end{equation}
where $\lambda$ is the wavelength of the carrier, and $L$ is the distance between the adjacent array elements at the BS. 
%For notational convenience, we will drop the subscript $j$ when the context allows.

%In practice, the available data are uniform time-samples of $\bm{y}(t)$ in the form of 
%\begin{equation}\label{eq:signal-samples}
%\bm{y}_{j}\left(m\right) = \sum_{k=1}^{K} \gamma_{j,k} %x\left(m/f_s-\tau_{j,k}\right) \bm{a}\left(\theta_{j,k}\right), \; m=0,1,\cdots,M-1,
%\end{equation}
%where $f_s$ is the sampling frequency, $M$ is the number of time instances for taking samples. 

%\begin{figure}
  %  \centering
  %  \includegraphics[width=0.35\textwidth]{Figures/DOA_v1.eps}
  %   \vspace{0.1cm}
  %   \caption{DoA of  LoS path  and  NLoS  paths }
   % \label{fig:DOA}
    
%\end{figure}

For the purpose of localisation, it is very important to note the different impacts of LoS path and NLoS paths. Denote the location of the MS by $\bm{l}_t = [l_t^x,l_t^y]^T$ where $l_t^x$ and $l_t^y$ are the horizontal and vertical coordinates, respectively. Assume that there is a LoS path between the MS and one BS located at $\bm{l}_{BS} = [l_{BS}^x, l_{BS}^y]^T$. Without loss of generality, assume that the position of the BS $\bm{l}_{BS}$ is given and the directions of the uniform arrays  are aligned with the horizontal axis. The ToA and DoA information is given by 

\begin{equation} \label{eq:LoS-tau}
\tau\left(\bm{l}_{t}\right) = {\Vert\bm{l}_{t}-\bm{l}_{BS}\Vert_{2}}/{c},
\end{equation}
\begin{equation} \label{eq:Los-theta}
\theta\left(\bm{l}_{t}\right) = %\mathrm{sign}\left(l_{t}^{y}-l_{BS}^{y}\right)
\mathrm{atan}({l_{t}^{x}-l_{BS}^{x}})/({l_{t}^{y}-l_{BS}^{y}}),
\end{equation}
respectively, where $c$ is the speed of light. There are two unknown variables in the MS location $l_t^x$ and $l_t^y$. They can be resolved from the ToA and DoA information defined in \eqref{eq:LoS-tau} and \eqref{eq:Los-theta}. 

The information provided by an NLoS path is very different. Let $\bm{l}_s = [l_s^x,l_s^y]^T$ be the location of a scatter, which is unknown to the BS. Then the ToA  and DoA   information is given by 
\begin{equation} \label{eq:NLoS-tau}
\tau\left(\bm{l}_{t},\bm{l}_{s}\right) = (\Vert\bm{l}_{t}-\bm{l}_{s}\Vert_{2} + \Vert\bm{l}_{s}-\bm{l}_{BS}\Vert_{2})/c,
\end{equation}
\begin{equation} \label{eq:NLos-theta}
\theta\left(\bm{l}_{t},\bm{l}_{s}\right) = %\mathrm{sign}\left(l_{s}^{y}-l_{BS}^{y}\right)
\mathrm{atan}( {l_{s}^{x}-l_{BS}^{x}} )/({l_{s}^{y}-l_{BS}^{y}}),
\end{equation}
respectively. Note that in this case ToA $\tau\left(\bm{l}_{t},\bm{l}_{s}\right)$ depends on both $\bm{l}_t$ and  $\bm{l}_s$ but DoA $\theta \left(\bm{l}_{t},\bm{l}_{s}\right)$ only depends on $\bm{l}_s$.

Due to their different impacts, LoS and NLoS paths are conventionally treated separately. The typical approach is to separate LoS path from NLoS paths and then use LoS path for localisation. In the case that there always exists a LoS path from MS to BS, the identification of LoS path can be achieved based on the fact that the LoS path experiences the minimum delay. However, in practice, there is no guarantee that a LoS path exists from the MS to a particular BS. The minimum delay path may correspond to a NLoS path. See \cite{borras1998decision,Markov_4,Markov_42,cong2001non,riba2004non, chan2006time,wylie1996non} for other more sophisticated methods to identifying LoS and NLoS paths. Note that mistakes in identification cannot be avoided completely and they lead to localisation performance deterioration. 

There are also efforts in the literature to perform localisation without distinguishing LoS/NLoS paths \cite{seow2008non,zhang2009combined,miao2007positioning}. The basic intuition is as follows. Based on the ToA/DoA estimates for each path, a geometric area of the possible locations of the MS can be inferred. If the intersection of all these geometric areas is a single point, then the MS must locate there. This approach suffers from the difficulty to infer the exact geometric areas for NLoS paths and rapid performance deterioration as noise increase. 

 %... (save some space ) ... (save some space ) ... (save some space ) ... (save some space ) ... (save some space ) ... (save some space ) ... (save some space ) ... (save some space )... (save some space ) ... (save some space ) ... (save some space ) ... (save some space )

\section{Our Solution \label{sec:Solution}}

\subsection{The Virtual Scatter} \label{subsec:VirtualScatter} 
\noindent
As detailed in Section \ref{sec:Problem}, the impacts of LoS and NLoS paths are different. The localisation process may fail if an NLoS path is mistakenly identified as a LoS path or vice versa. In the literature, many efforts were put to distinguish them. 

\begin{figure}
    \hspace*{\fill}
	\subfloat[\label{fig:sep_vir_demo}]{\includegraphics[width=0.12\textwidth]{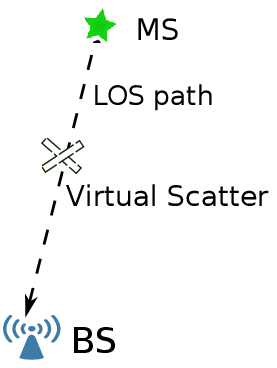}}\hfill{}\subfloat[\label{fig:sJoint_vir_demo}]{\includegraphics[width=0.15\textwidth]{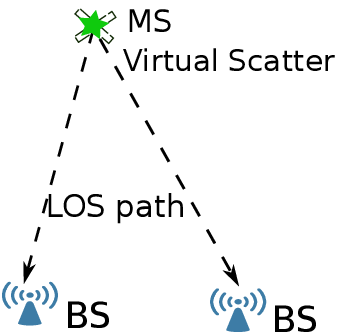}}
	\hspace*{\fill}
	\caption{(a) When there is only one LoS path existing in the localisation problem, the location of the virtual scatter can be arbitrary on the line segment of that particular LoS path. (b) When multiple LoS paths are used for localisation, the most parsimonious solution (in number of scatters) is unique where a single virtual scatter is created at the intersection of all LoS paths. \label{fig:The-virtual-scatter} }
\end{figure}

A key element of our approach is to introduce the concept of virtual scatter so that there is no need to distinguish LoS and NLoS paths. This is achieved by introducing virtual scatter in the LoS path between the MS and the BS. See Fig. \ref{fig:The-virtual-scatter}(a) for an illustration. With the virtual scatter, the ToA and DoA information of both LoS and NLoS paths can be consistently presented by \eqref{eq:NLoS-tau} and \eqref{eq:NLos-theta}, respectively. The mathematical models for LoS and NLoS paths are therefore united. This unified model eventually leads to a simplified and tractable formulation in \eqref{eq:Y-linear-combination} without handling LoS and NLoS paths separately.

It is worth to comment more on the virtual scatter. When there is only one LoS path existing in the localisation problem, the location of the virtual scatter can be arbitrary on the line segment of that particular LoS path. This arbitrariness will not affect the localisation performance as the ultimate goal is to locate the MS. When there are multiple LoS paths in the localisation problem, the most parsimonious solution (in number of scatters) is unique where a single virtual scatter is created at the intersection of all LoS paths, i.e., the location of the MS. This idea is illustrated in Fig. \ref{fig:The-virtual-scatter}(b). A similar idea was briefly mentioned in \cite{gear2019maximum} but not carefully engineered into the localisation algorithm. 

%\todowd{A reminder on how much details we should talk about the non-uniqueness.}

\subsection{Direct Localisation as Linear De-Mixing \label{subsec:DirectLocalisation}}
\noindent
There are different ways to perform localisation. Many works in the literature \cite{seow2008non,zhang2009combined,miao2007positioning} take the approach of first estimating ToA and DoA information and then using the estimated ToA and DoA for localisation. As the same scatter creates different paths and gives rise to different ToA and DoA information different BSs. Data association is needed. However, data association is known to be NP-hard and error propagation is inevitable. 

Another way is to embed the geometry information into an optimisation formulation, also known as direct localisation \cite{VTC2008,shahmansoori2017position,garcia2017direct,gear2019maximum}. Rather than estimating ToA $\tau$  and DoA $\theta$, the unknown variables are chosen as the location of MS $\bm{l}_t$ and that of scatters $\bm{l}_s$. In other words, the signal model is the `direct' map from $\left(\bm{l}_{t},\bm{l}_{s}\right)$ to received signal $\bm{y}_j(t)$.  This strategy avoids data association. 

It is challenging to solve the inverse problem based on the nonlinear map from $\left(\bm{l}_{t},\bm{l}_{s}\right)$ to $\bm{y}$. There is a huge literature on estimating the intermediate variables ToA $\tau$ and DoA $\theta$, because of nice  structures of the associated problem. In direct localisation, these structures cannot be directly explored. To avoid directly handling the nonlinear map from $\left(\bm{l}_{t},\bm{l}_{s}\right)$ to $\bm{y}$, we formulate the forward model into a linear form. Consider the map from locations to the $j$-th data matrix
\begin{align} \label{eq:mapping-location-data}
\left(\bm{l}_{t},\bm{l}_{s}\right)\in\mathbb{R}^{2}\times\mathbb{R}^{2} & \mapsto\bm{B}_j\in\mathbb{C}^{N_{R}\times N},
\end{align}
where $\bm{B}_j=\bm{a}\left(\theta_j\right)\left[x(0-\tau_j),\cdots,x(N-1-\tau_j)\right]$.
Then the received data $\bm{Y}_j = [\bm{y}_j(0),\cdots,\bm{y}_j(N-1)]\in \mathbb{C}^{N_R \times N}$ is a linear combination of the responses of scatters 
\begin{equation} \label{eq:Y-linear-combination}
	\bm{Y}_j = \sum_k \gamma_k \bm{B}_j (\bm{l}_t,\bm{l}_{s,k}).
\end{equation}
Localisation problem is then reduced to a problem of de-mixing a linear combination of atoms from the set of $\bm{B}_j (\bm{l}_t,\bm{l}_{s})$. 

\subsection{Super-resolution formulation \label{subsec:super-resolution}}
\noindent
There are two approaches to solve the demixing problem in \eqref{eq:Y-linear-combination}. One approach relies on discretising the space ${(\bm{l}_t,\bm{l}_s)}$. By such a discretisation, one obtains a final size dictionary of $\bm{B} (\bm{l}_t,\bm{l}_{s})$. Suppose that the number of scatters is much smaller than the number of discrete grid points, the de-mixing problem \eqref{eq:Y-linear-combination} can be solved by sparse recovery techniques \cite{dai2009subspace,pati1993orthogonal}. 

Our approach is to estimate $(\bm{l}_t,\bm{l}_{s})$ as continuous parameters. The dictionary of  $\bm{B} (\bm{l}_t,\bm{l}_{s})$ contains uncountable many atoms. One technique to solve such an inverse problem is super-resolution. 

There are many reasons working on a continuous parameter space rather than the discrete counterpart. First, discretisation leads to the leakage effect when  the ground-truth is off-the-grid. The signal generated from an off-the-grid $(\bm{l}_t,\bm{l}_{s})$ typically cannot be well approximated by signals from a small number of grid points. It has been observed that off-the-grid points typically result in approximation error that is proportional to signals themselves, and deteriorate the estimation performance. Second, a finer grid may mitigate the off-the-grid leakage but may result in large computational cost. Each tuple $(\bm{l}_t,\bm{l}_{s})$ has four variables $l_t^x,l_t^y,l_s^x,l_s^y$. An $N_d$-point grid for each variable results in $N_d^4$ many grid points. By contrast, super-resolution approach works on the continuous parameter space, avoids the leakage effect, and its complexity mainly depends on the number of data samples $N$, which can be many orders of magnitude less than the number of grid points $N_d^4$.

% \subsection{Algorithm Details \label{subsec:algorithm}}

In the super-resolution framework, the locations are encoded using atomic measures 
\begin{equation}
\mu_{j}=\sum_{k=1}^{K}\gamma_{j,k}\delta_{(\bm{l}_{t},\bm{l}_{s})_{k}},\; j=1,...,J,
\label{atomic measure}
\end{equation}
where $\delta_{(\bm{l}_{t},\bm{l}_{s})}$ 
is Dirac function at $\left(\bm{l}_{t},\bm{l}_{s}\right)$. The subscript $j$ emphasises that the attenuation coefficients $\gamma_{j}$ may be different for different BSs. The received signal at the $j$-th BS is then 
\begin{equation}
\bm{Y}_j = \bm{\varPsi}_{j} \mu_{j} := \int_{(\bm{l}_{t},\bm{l}_{s})\in\mathbb{R}^{2}\times\mathbb{R}^{2}}\bm{B}_{j}(\bm{l}_t,\bm{l}_{s})\mathrm{d}\mu_{j}\left(\bm{l}_{t},\bm{l}_{s}\right),
\end{equation}
where  $\bm{\varPsi}_{j}$ is a linear operator mapping the atomic measure $\mu_j$ to $\bm{Y}_j \in \mathbb{C}^{N_R \times R}$, and $\bm{B}_j(\bm{l}_t,\bm{l}_{s})=\bm{a}( \theta_j )\left[x(0-\tau_j),\cdots,x(N-1-\tau_j)\right]$.

A total variation (TV) norm and a group total variation (GTV) norm are defined for atomic measures to promote certain sparse structures of locations. In practice, there are typically a small number of MS and scatter pairs compared with the signal dimension $N_R\times N$. The total variation norm is defined to promote the sparsity of the MS and scatter pairs: 
\begin{equation}
\left\Vert \mu_j \right\Vert _{\mathrm{TV}}:= \underset{|\alpha (\bm{l}_t,\bm{l}_{s})| \le 1}{\sup} {\rm Re} \int \alpha(\bm{l}_{t},\bm{l}_{s}) \mathrm{d} \mu_j (\bm{l}_t,\bm{l}_{s}) 
=\sum_{k}|\gamma_{j,k}|.\label{eq:TV}
\end{equation} 
It is an analogy to $\ell_{1}$-norm for finite dimensional vector space \cite{candes2014towards}. 
Note that for given MS and scatter pair located at $(\bm{l}_t,\bm{l}_s)$, it likely generates signals received at multiple BSs. Motivated by the group sparsity concept commonly used in the literature of compressed sensing, the GTV norm is defined as  
\begin{align}
\left\Vert \bm{\mu}\right\Vert _{\mathrm{GTV}} & :=\underset{\Vert\bm{\alpha}(\bm{l}_{t},\bm{l}_{s})\Vert_{2}\le1}{\sup}\;\sum_{j=1}^{J}\mathrm{Re} \int\alpha_{j}(\bm{l}_{t},\bm{l}_{s}){\rm d}\mu_{j}(\bm{l}_{t},\bm{l}_{s}) \nonumber \\
& =\sum_{k}\left\Vert \bm{\gamma}_{k}\right\Vert _{2},\label{eq:GTV-def}
\end{align}
where $\bm{\mu} = [\mu_1,\cdots, \mu_J]^T$, $\bm{\alpha}(\bm{l}_{t},\bm{l}_{s}) = [\cdots, \alpha_j(\bm{l}_{t},\bm{l}_{s}), \cdots]^T$, and $\bm{\gamma}_k = [\gamma_{1,k},\cdots,\gamma_{J,k}]^T$.
Clearly it is an analogy to $\ell_{2,  1}$ norm  \cite{fernandez2016super}. 
 
With above definitions, the localisation problem in the noise free case can be then formulated as 
\begin{equation}
\underset{\bm{\mu}}{\min} \; \sum_{j=1}^{J} \left\Vert \bm{\mu}_{j}\right\Vert _{\mathrm{TV}} + \lambda \left\Vert \bm{\mu}\right\Vert _{\mathrm{GTV}}\;{\rm s.t.}\;\bm{Y}_{j}=\bm{\varPsi}_{j}\mu_{j},\;\forall j\in[J], \label{eq:TVMINI}
\end{equation}
where $\lambda>0$ is a regularisation constant, and $[J]:=\{1,\cdots,J\}$. For the noisy case, either replace the equality  $\bm{Y}_{j}=\bm{\varPsi}_{j}\mu_{j}$ with $\Vert \bm{Y}_{j} - \bm{\varPsi}_{j}\mu_{j} \Vert_F \le \epsilon_j$ when where the noise power is known at most $\epsilon_j > 0$, or turn the constrained optimisation \eqref{eq:TVMINI} into a Lasso type unconstrained optimisation. Note that in order to ensure the convexity of \eqref{eq:TVMINI}, we do not restrict the number of  possible $\bm{l}_t$ to one. Nevertheless, all trials in our simulations return a single MS location.

%Note that the inverse problem under the TV norm and GTV norm involves parameter from infinite dimensional space. Some super-resolution techniques as \cite{candes2014towards,tang2013compressed,fernandez2016super}, it shows that line spectral estimation problem can be formulated as semi-definite programming (SDP) with finite many variables and constraints. However, this framework cannot be directly extended to our model, because the received signal is not the superposition of simple exponential form. Actually, our model is  mixtures with a mapping from the locations $ \left(\bm{l}_{t},\bm{l}_{s,k}\right)$ to the corresponding  data matrix $\bm{B}_{j}(\bm{l}_t,\bm{l}_{s})$ with irregular trigonometric polynomial form. 

\subsection{Solving the super-resolution problem}
\noindent
It is highly non-trivial to solve the convex optimisation problem \eqref{eq:TVMINI}. The biggest hurdle is that \eqref{eq:TVMINI} is a sparse encoding using a dictionary with infinite many atoms, hence an optimisation problem with infinite many unknown variables. To overcome the technical difficulty, the method alternating descent conditional gradient (ADCG) \cite{boyd2017alternating} is adopted and slightly modified here.  See Algorithm \ref{alg:Super-resolution}  for a highly level description. It is also noteworthy that the gradient computations in Algorithm \ref{alg:Super-resolution} are related to the specific waveform $x(t)$ used in \eqref{eq:signal-model} and non-trivial. Details are omitted due to space constraint and will be presented in the journal paper.

\begin{algorithm}
\textbf{\small{}Input}{\small{} : $\;$Candidate set $\mathcal{S}_{0}=\emptyset$,$\mathrm{threshold}$, $\bm{\mathrm{Y}_j},j=1,...,J$. ,
}{\small\par}

\textbf{\small{}Output}{\small{} :$\;$Locations of MS
and scatters $:\mathcal{S}_{K}$. }{\small\par}

\textbf{\small{}Iteration}{\small{} $k$}{\small\par}
\begin{itemize}
\item {\small{}Compute gradient of loss : $\bm{g}_{j,k} = \nabla\left\Vert \bm{r}_{j,k}\right\Vert _{2}^{2},$
where $\bm{r}_{j,k}=\boldsymbol{\varPsi}_{j}\mu_{k-1}-\bm{Y}_{j}$}{\small\par}
\item {\small{}Compute next source :}\\
$\left(\bm{l}_{t},\bm{l}_{s}\right)_k=\mathrm{arg\,min}\sum_{j=1}^{J}\left\langle \boldsymbol{B}_{j}\left(\bm{l}_{t},\bm{l}_{s}\right)_k,\boldsymbol{g}_{j,k}\right\rangle$ .
\item {\small{}Update the candidate set  : $\mathcal{S}_{k}=\mathcal{S}_{k-1}\cup\left\{\bm{l}_{t},\bm{l}_{s}\right\}_k$. }{\small\par}
%\item {\small{}Update the candidate set of MSs : $\bm{MS}_{k}=\left[\bm{MS}_{k-1},\bm{l}_{t,k}\right]$.}{\small\par}
\item {\small{}Coordinate descent}{\small\par}
\begin{itemize}
\item {\small{}Compute weights :}\\
{\small{} $\bm{\gamma}=\mathrm{arg\,min}\sum_{j}^{J}\left\Vert \bm{Y}_{j}-\boldsymbol{B}_{j}\left(\mathcal{S}_{k}\right)\bm{\gamma}_{:,j}\right\Vert _{2}^{2}+\lambda_{1}\sum_{j}^{J}\left\Vert \bm{\gamma}_{:,j}\right\Vert _{1}+\lambda_{2}\left\Vert \bm{\gamma}\right\Vert _{2,1}$}{\small\par}
\item {\small{}Prune support :}\\
{\small{} $\ensuremath{\mathcal{S}_{k}=\left\{ \left[\boldsymbol{l}_{t,i},\boldsymbol{l}_{s,i}\right]\in\mathcal{S}_{k}|\underset{i}{\mathrm{min}}\,\left\Vert \bm{\gamma}_{i,1:J}\right\Vert _{2}>\mathrm{threshold}\right\} } $}{\small\par}

\item {\small{}Locally improve support :\\
$ \mathcal{S}_{k}=\mathrm{gradient}\textrm{\_}\mathrm{descent}\left(\big(\left(\bm{l}_{s},\bm{l}_{t}\right),\bm{\gamma}(\left\{\bm{l}_{s},\bm{l}_{t}\right\})\big):\left\{\bm{l}_{s},\bm{l}_{t}\right\}\in\mathcal{S}_{k}\right)$}{\small\par}
\end{itemize}
\end{itemize}
{\small{}}{\small\par}

{\small{}\caption{ADCG in multiple cooperative BSs system \label{alg:Super-resolution}}
}{\small\par}
\end{algorithm}

\section{Numerical Simulations}

\subsection{Simulation Setup}
\noindent
The transmitted signal can be chosen for different purpose.
In this paper, we take one block of orthogonal frequency division
multiplexing (OFDM) as an example, since it is used to combat multi-path
fading and achieve high spectral efficiency. Suppose the transmitted signal from MS is 
\begin{equation}
x\left(t\right)=\sum_{n=0}^{N-1}s\left(n\right)e^{i2\pi n\triangle f t},
\end{equation}
where  $s\left(n\right)$ is the data symbol, $\triangle f=1/T$ is the sub-carrier frequency, and $T$ is
the duration of each block.

By taking the Fourier transform on the received signal
at the BS $j$ , the observation at $n$-th sub-carrier can be written
as 
\begin{equation}
\bm{Y}_{j}\left(n\right)=\int_{t=0}^{T}e^{-i2\pi n\triangle ft}\bm{y}_{j}\left(t\right)\mathrm{dt}.
\end{equation}
In practice, the available data are uniform time-samples of $\bm{y}(t)$. In this case, the integrals can be replaced by summation of  time instances. The matrix form of measurement at receiver $j$ is
\begin{align}
\bm{Y}_{j} =\sum_{k=1}^{K}\gamma_{j,k}\bm{B}_{j}\textrm{\ensuremath{\left(\theta_{j,k},\tau_{k}\right)}} =
\sum_{k=1}^{K}\gamma_{j,k}\textrm{\ensuremath{\bm{a}\left(\theta_{j,k} \right)\bm{b}^{T}\left(\tau_{j,k}\right)}},
\nonumber 
\end{align}
where the vector $\bm{b}\left(\tau_{j,k}\right)\in\mathrm{C}^{N}$
associated with time delay for $k$-th propagation path is expressed
as

\begin{align}
\bm{b}\left(\tau_{j,k}\right) & =\left[s(0),\cdots,s(N-1)\left(e^{-i2\pi\triangle f\tau_{j,k}\left(\bm{l}_{t},\bm{l}_{s,k}\right)\left(N-1\right)}\right)\right]^{\mathrm{T}}.\nonumber 
\end{align}

In this section, we apply the super-resolution  based approach to estimate the
localisation of MS under the 4 cooperative base stations which  are horizontally
located at $\left(0,0\right)\mathrm{km},\left(0,1\right)\mathrm{km},\left(1,0\right)\mathrm{km}$,
$\left(1,1\right)\mathrm{km}$. We consider the scenario that the MS, scatters and BSs are located
in a $1\mathrm{km}\times1\mathrm{km}$ area. 
In simulation, the transmitted OFDM signal  contains  $N=32$ sub-carrier frequencies,
the sub-carrier frequency spacing is $\triangle f=10kHz$ , the speed
of light is $c=3\times10^{8}\mathrm{m/s}$, carrier frequency $f_{c}=2GHz$
.

\subsection{Results and Analysis }
\begin{figure}
    \hspace*{\fill}
	\subfloat[\label{fig:Results}]{\includegraphics[width=0.26\textwidth]{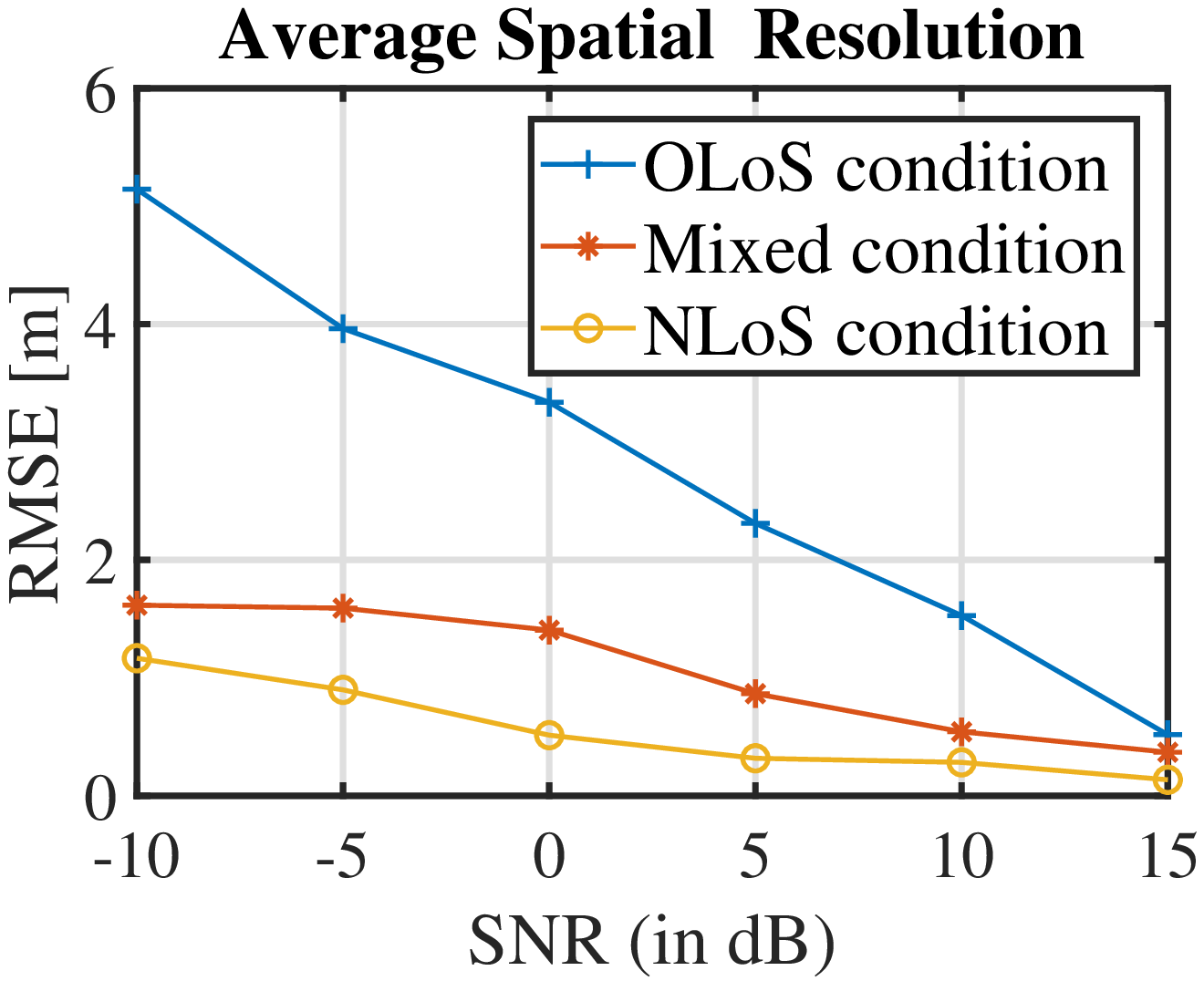}}\hfill{}\subfloat[\label{fig:One_case}]{\includegraphics[width=0.215\textwidth]{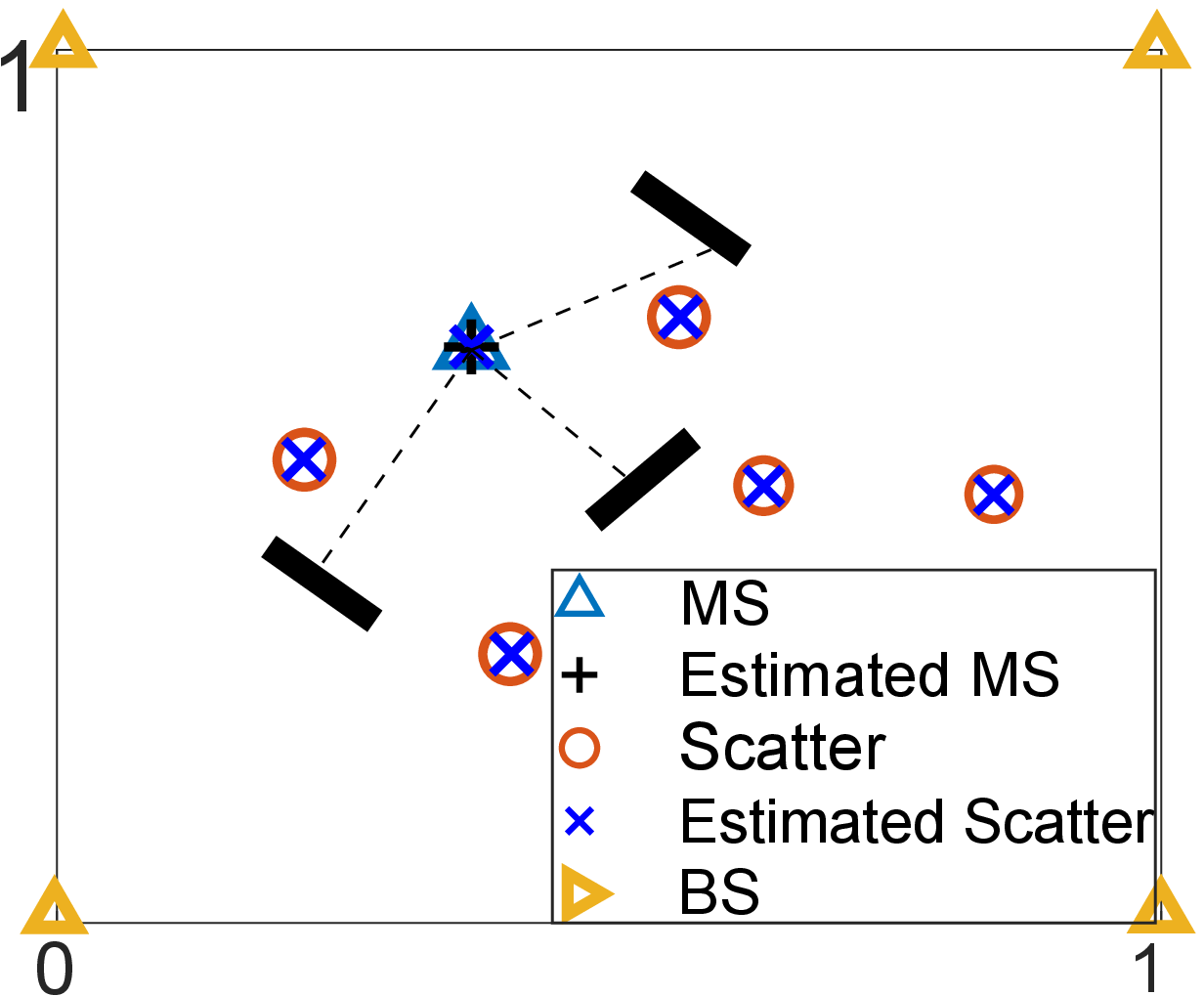}}
	\hspace*{\fill}
	\caption{(a) Simulation results based on Monte Carlo. (b) One trial of mixed condition with SNR 
	$= 0$ dB \label{Final_comp}. }
\end{figure}

\noindent
Fig. \ref{Final_comp}(a) depicts the root mean square error (RMSE) of proposed super-resolved localisation scheme under unknown propagation conditions. The RMSE illustrates the estimated spatial resolution of both MS and scatters, and it is defined as $\ensuremath{\mathrm{RMSE}=({\sum_{k=1}^{K}\parallel\boldsymbol{l}_{s,k}-\widetilde{\boldsymbol{l}}_{s,k}\parallel_{2}+\parallel\boldsymbol{l}_{t}-\widetilde{\boldsymbol{l}}_{t}\parallel_{2}})/({K+1}})$, where $\widetilde{\boldsymbol{l}}_{s,k}$, $\widetilde{\boldsymbol{l}}_{t}$ are estimated locations of  scatters ${\boldsymbol{l}}_{s,k}$ and MS ${\boldsymbol{l}}_{t}$ respectively. The output of the proposed approach contains one MS and $K$ scatters, the correspondence of estimated MS and ground truth is easily determined. The correspondence of scatter can be determined by finding ground truth which is  nearest to estimated scatter. For each RMSE, we run a total of 300 Monte Carlo trials, and randomly generate locations of MS and scatters. In the simulation, the OLoS condition refers to all the BSs have the OLoS condition, and the same concept can be applied to OLoS condition. The mixed condition models the general environment that each BS faces either OLoS condition or NLoS condition. We can see that the performance of proposed algorithm depends on the siganl-to-noise ratio (SNR) and the number of propagation paths. For each path, we add extra Additive White Gaussian Noise (AWGN) to noise environment. The simulation shows that proposed approach is robust to noise. It is observed that the localisation of MS and scatters
is of high precision. Even in the worst case  (SNR = -10dB), the spatial resolution in NLoS, mixed and OLoS condition are approximately 1.17$\mathrm{m}$ 1.61 $\mathrm{m}$ and 5.14$\mathrm{m}$. The proposed algorithm has better performance in NLoS condition. This is due to the NLoS condition containing both LoS paths and scattered paths, which has more measurements than mixed and OLoS condition. 

Fig. \ref{Final_comp}(b) illustrates one trial of mixed condition with the SNR = 0 dB. In this case, one BS faces to NLoS condition, and the rest of BSs are OLoS condition. Without any prior knowledge of the propagation conditions, the proposed unified localisation approach can be applied in
mixed propagation environments. The approach can not only estimate
the localisation of the MS to a high precision, but also determine locations of scatters.

%\begin{figure}
  %  \centering
  %  \includegraphics[width=0.3\textwidth]{Figures/RMSE_SNR_NLOS_OLOS.eps}
  %   \vspace{0.1cm}
  %   \caption{Average Estimated Spatial Resolution of MS and Scatters  }
  %  \label{fig:SIM_NLOS}
    
%\end{figure}

\newpage

\bibliographystyle{IEEEtran}
\bibliography{Ref,Intro}

\begin{thebibliography}{10}
\providecommand{\url}[1]{#1}
\csname url@samestyle\endcsname
\providecommand{\newblock}{\relax}
\providecommand{\bibinfo}[2]{#2}
\providecommand{\BIBentrySTDinterwordspacing}{\spaceskip=0pt\relax}
\providecommand{\BIBentryALTinterwordstretchfactor}{4}
\providecommand{\BIBentryALTinterwordspacing}{\spaceskip=\fontdimen2\font plus
\BIBentryALTinterwordstretchfactor\fontdimen3\font minus
  \fontdimen4\font\relax}
\providecommand{\BIBforeignlanguage}[2]{{%
\expandafter\ifx\csname l@#1\endcsname\relax
\typeout{** WARNING: IEEEtran.bst: No hyphenation pattern has been}%
\typeout{** loaded for the language `#1'. Using the pattern for}%
\typeout{** the default language instead.}%
\else
\language=\csname l@#1\endcsname
\fi
#2}}
\providecommand{\BIBdecl}{\relax}
\BIBdecl

\bibitem{VTC2008}
K.~Papakonstantinou and D.~Slock, ``Direct location estimation using
  single-bounce {NLOS} time-varying channel models,'' in \emph{2008 IEEE 68th
  Vehicular Technology Conference}, Sept 2008, pp. 1--5.

\bibitem{shahmansoori2017position}
A.~{Shahmansoori}, G.~E. {Garcia}, G.~{Destino}, G.~{Seco-Granados}, and
  H.~{Wymeersch}, ``Position and orientation estimation through millimeter-wave
  {MIMO} in 5g systems,'' \emph{IEEE Transactions on Wireless Communications},
  vol.~17, no.~3, pp. 1822--1835, March 2018.

\bibitem{garcia2017direct}
N.~Garcia, H.~Wymeersch, E.~G. Larsson, A.~M. Haimovich, and M.~Coulon,
  ``Direct localization for massive {MIMO},'' \emph{IEEE Transactions on Signal
  Processing}, vol.~65, no.~10, pp. 2475--2487, 2017.

\bibitem{gear2019maximum}
B.~{Gear}, E.~{Mellios}, A.~{Nix}, and J.~{McGeehan}, ``A maximum likelihood
  location estimator for non-line of sight geolocation of radio emitters,'' in
  \emph{2019 13th European Conference on Antennas and Propagation (EuCAP)},
  March 2019, pp. 1--5.

\bibitem{borras1998decision}
J.~Borras, P.~Hatrack, and N.~B. Mandayam, ``Decision theoretic framework for
  {NLOS} identification,'' in \emph{VTC'98. 48th IEEE Vehicular Technology
  Conference. Pathway to Global Wireless Revolution (Cat. No. 98CH36151)},
  vol.~2.\hskip 1em plus 0.5em minus 0.4em\relax IEEE, 1998, pp. 1583--1587.

\bibitem{Markov_4}
L.~Chen and L.~Wu, ``Mobile positioning in mixed {LOS/NLOS} conditions using
  modified {EKF} banks and data fusion method,'' \emph{IEICE Transactions on
  Communications}, vol.~92, no.~4, pp. 1318--1325, 2009.

\bibitem{Markov_42}
C.~Morelli, M.~Nicoli, V.~Rampa, and U.~Spagnolini, ``Hidden {Markov} models
  for radio localization in mixed {LOS/NLOS} conditions,'' \emph{IEEE
  Transactions on Signal Processing}, vol.~55, no.~4, pp. 1525--1542, April
  2007.

\bibitem{cong2001non}
L.~Cong and W.~Zhuang, ``Non-line-of-sight error mitigation in {TDOA} mobile
  location,'' in \emph{GLOBECOM'01. IEEE Global Telecommunications Conference
  (Cat. No. 01CH37270)}, vol.~1.\hskip 1em plus 0.5em minus 0.4em\relax IEEE,
  2001, pp. 680--684.

\bibitem{riba2004non}
J.~Riba and A.~Urruela, ``A non-line-of-sight mitigation technique based on
  {ML}-detection,'' in \emph{2004 IEEE International Conference on Acoustics,
  Speech, and Signal Processing}, vol.~2.\hskip 1em plus 0.5em minus
  0.4em\relax IEEE, 2004, pp. ii--153.

\bibitem{chan2006time}
Y.-T. Chan, W.-Y. Tsui, H.-C. So, and P.-c. Ching, ``Time-of-arrival based
  localization under {NLOS} conditions,'' \emph{IEEE Transactions on Vehicular
  Technology}, vol.~55, no.~1, pp. 17--24, 2006.

\bibitem{wylie1996non}
M.~P. {Wylie} and J.~{Holtzman}, ``The non-line of sight problem in mobile
  location estimation,'' in \emph{Proceedings of ICUPC - 5th International
  Conference on Universal Personal Communications}, vol.~2, Oct 1996, pp.
  827--831 vol.2.

\bibitem{seow2008non}
C.~K. Seow and S.~Y. Tan, ``Non-line-of-sight localization in multipath
  environments,'' \emph{IEEE Transactions on Mobile Computing}, vol.~7, no.~5,
  pp. 647--660, 2008.

\bibitem{zhang2009combined}
V.~Y. Zhang and A.~K.-s. Wong, ``Combined {AOA} and {TOA} {NLOS} localization
  with nonlinear programming in severe multipath environments,'' in \emph{2009
  IEEE Wireless Communications and Networking Conference}.\hskip 1em plus 0.5em
  minus 0.4em\relax IEEE, 2009, pp. 1--6.

\bibitem{miao2007positioning}
H.~Miao, K.~Yu, and M.~J. Juntti, ``Positioning for {NLOS} propagation:
  Algorithm derivations and {C}ramer--{R}ao bounds,'' \emph{IEEE Transactions
  on Vehicular Technology}, vol.~56, no.~5, pp. 2568--2580, 2007.

\bibitem{yin2013and}
F.~Yin, C.~Fritsche, F.~Gustafsson, and A.~M. Zoubir, ``{EM}-and {JMAP-ML}
  based joint estimation algorithms for robust wireless geolocation in mixed
  {LOS/NLOS} environments,'' \emph{IEEE Transactions on Signal Processing},
  vol.~62, no.~1, pp. 168--182, 2013.

\bibitem{dai2009subspace}
W.~Dai and O.~Milenkovic, ``Subspace pursuit for compressive sensing signal
  reconstruction,'' \emph{IEEE transactions on Information Theory}, vol.~55,
  no.~5, pp. 2230--2249, 2009.

\bibitem{pati1993orthogonal}
Y.~C. {Pati}, R.~{Rezaiifar}, and P.~S. {Krishnaprasad}, ``Orthogonal matching
  pursuit: recursive function approximation with applications to wavelet
  decomposition,'' in \emph{Proceedings of 27th Asilomar Conference on Signals,
  Systems and Computers}, Nov 1993, pp. 40--44 vol.1.

\bibitem{candes2014towards}
E.~J. Cand{\`e}s and C.~Fernandez-Granda, ``Towards a mathematical theory of
  super-resolution,'' \emph{Communications on Pure and Applied Mathematics},
  vol.~67, no.~6, pp. 906--956, 2014.

\bibitem{fernandez2016super}
C.~Fernandez-Granda, ``Super-resolution of point sources via convex
  programming,'' \emph{Information and Inference: A Journal of the IMA},
  vol.~5, no.~3, pp. 251--303, 2016.

\bibitem{boyd2017alternating}
N.~Boyd, G.~Schiebinger, and B.~Recht, ``The alternating descent conditional
  gradient method for sparse inverse problems,'' \emph{SIAM Journal on
  Optimization}, vol.~27, no.~2, pp. 616--639, 2017.

\end{thebibliography}

\end{document}